\documentclass[onecolumn,showpacs,preprintnumbers,amsmath,amssymb,pra]{revtex4}
\usepackage{txfonts}
\usepackage{graphicx}
\usepackage{dcolumn}
\usepackage{bm}
\begin{document}

\title{single-photon frequency conversion and multi-mode entanglement via constructive interference on Sagnac Loop}
\author{Anshou Zheng} \email{zaszas1_1@126.com}
\author{Guangyong Zhang}
\author{Liangwei Gui}
\affiliation{School of Mathematics and Physics, China University of Geosciences, Wuhan, 430074, P. R. China}

\author{Jibing Liu}\email{liu0328@foxmail.com}

\affiliation{Hubei Key Laboratory of Pollutant Analysis and Reuse Technology and Department of Physics,
Hubei Normal University, Huangshi, 435002, People¡¯s Republic of China}

\date{\today}

\begin{abstract}
Based on constructive interference in Sagnac waveguide loop,
an efficient scheme is proposed for selective frequency conversion and multifrequency
modes $W$ entanglement via input-output formalism.
We can adjust the probability amplitudes of output photons by choosing
parameter values properly. The tunable probability amplitude will
lead to the generation of output photon with a selectable frequency
 and $W$ photonic entanglement of different frequencies modes in
  a wide range of parameter values.
 Our calculations show the present scheme is robust to the deviation of parameters and
 spontaneous decay.

\end{abstract}


\maketitle
\section{Introduction}
Quantum frequency conversion has been attracting attention as a potentially crucial
resource in interfacing photonic quantum information systems working at disparate
frequencies. Each kind of optical device has its own optimal frequencies which
minimize the loss. For example, photon with wavelength around $800$ nm is appropriate
for the coherence optical memories with highest efficiency \cite{Nc2-174}.
Recently, some works \cite{prl92-68,prl03-90,N05-437,Np10-4,prl10-105,Nc11-2,prl12-109}
 demonstrate that the quantum character of the light field,
such as entanglement and antibunching of light, is conserved during the process of
quantum frequency conversion. This popularizes the use of frequency
conversion in quantum information science, such as quantum repeaters,
quantum memories, quantum key distribution, etc. In addition,
quantum frequency conversion is widely employed in high-sensitivity
detection of optical signals \cite{pr62-127,Jap67-38} and single-photon
source. So far, a great variety of methods
\cite{prl92-68,prl03-90,N05-437,Np10-4,prl10-105,Nc11-2,prl12-109,pr62-127,Jap67-38,Ol29-1449,
Praz12-86,Jmo04-51,Ol05-30,Nc2-174,Oe12-4,Ol12-37,Ol10-35,Pra10-82,Sci12-338,Nc12-3,
Apl12-101,prl12-108,pra12-85,Prawu04-69,olwu04-29,d,wu1,wu11} are provided for
single-photon frequency conversion, like multiwave mixing
\cite{prl92-68,N05-437,Oe12-4,prl10-105,Ol12-37,Ol10-35,wu11,Pra10-82},
cavity opto-mechanics system \cite{Sci12-338,Nc12-3} and
single-photon adiabatic wavelength conversion \cite{Apl12-101},
and so on.
More recently, a novel theoretical scheme is proposed for
single-photon frequency conversion through constructive interference
in Sagnac loop \cite{prl12-108,pra12-85}. We extend the constructive
interference scheme to the realization of selective single-photon frequency conversion and
multifrequency photonic entanglement, i.e., $W$ entangled state \cite{w}
, which has many inequivalent classes and cannot be transformed
into each other under local operations and classical communication (LOCC) protocols.
Thus, it is a crucial resource in quantum information science.

The present scheme has several important features. 
(1) The frequency of the output photon is adjustable by stark shift \cite{e-p}
and we can also elect the frequency of the output photon by  adjusting parameters.
In addition, the present scheme is capable of generating
multifrequency photonic entanglement. (2) It does not require phase-matching on
the involved frequencies of the optical fields although the similar multi-$\Lambda$ model is employed in
multiwave mixing process \cite{olwu04-29,d,wu1,li}. Thus, we can choose the energy model widely for the present scheme.
Note that tunable and selective wavelength conversion is described using fibre four-wave mixing \cite{ee},
however the phase-matching condition is required.
(3) The same probability amplitudes of the output photons will lead to the $W$ entangled state of
photons with different frequencies. In principle, it can be extended to the $N$ mode
photonic entanglement by choosing energy model and parameters properly.

\section{The Model}
The model under consideration is illustrated in Fig. 1. It consists of a five-level atom coupling directly
to a Sagnac waveguide loop. An external photon can go into the waveguide loop
through a $50:50$ beam splitter. On the waveguide loop, a phase shifter
is used to adjusted the relative phase $(\Theta)$ between the clockwise
and counterclockwise propagating pulses.
 As shown in Fig. 1(b) the atom has a multi-$\Lambda$-type level
configuration, ground states $|0\rangle$ and $|e\rangle$, excited states $|1\rangle$, $|2\rangle$ and
$|3\rangle$. Three continuous-wave (cw) $\Omega_1$,
$\Omega_2$ and $\Omega_3$ are used, respectively, to driven resonantly the
atomic transitions $|1\rangle\Longleftrightarrow|e\rangle$,
$|2\rangle\Longleftrightarrow|e\rangle$, and $|3\rangle\Longleftrightarrow|e\rangle$.
Another three transitions $|0\rangle\Longleftrightarrow|1\rangle$,
$|0\rangle\Longleftrightarrow|2\rangle$ and $|0\rangle\Longleftrightarrow|3\rangle$
are driven by three quantum fields $g_1$, $g_2$ and $g_3$, respectively. In the
interaction picture, the interaction Hamiltonian of the coupled system can given
as ($\hbar=1$) \cite{prawu96-54,prad05-72,oejh11-19}:
\begin{align}
H_I=\int_{-\infty}^{+\infty}\sum_{j=1,2,3}^{k=L,R}[\Delta_ja_{jk}^\dagger(\omega) a_{jk}(\omega)
-g_ja_{jk}^\dagger(\omega)\sigma_{0j}-g_ja_{jk}(\omega)\sigma_{j0}]d\omega
-\sum_{j=1,2,3}\Omega_j(\sigma_{je}+\sigma_{ej}),
\end{align}
\begin{figure}[here]
\includegraphics[width=0.3\textwidth]{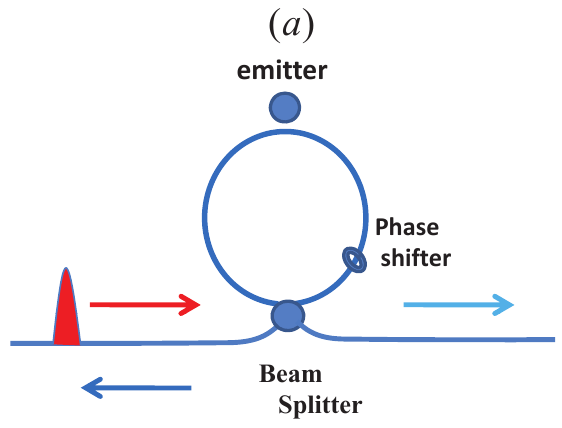}
\includegraphics[width=0.26\textwidth]{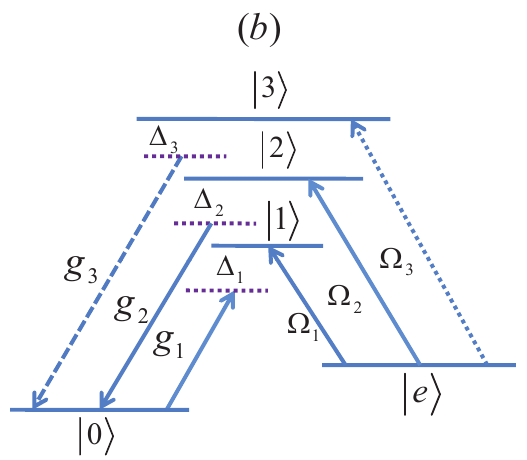}
\includegraphics[width=0.28\textwidth]{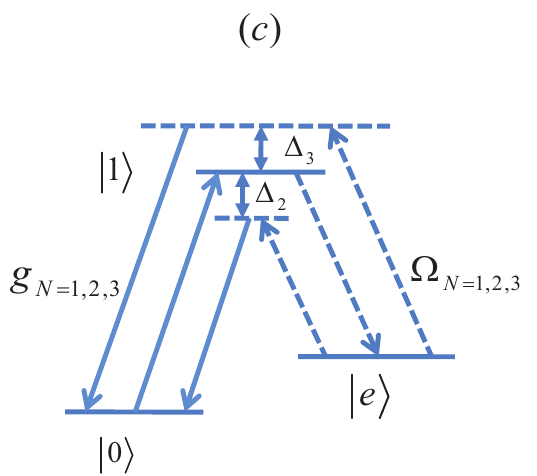}
\caption{\label{fig:fig4}(Color online)
(a) Schematic of optical system. Sagnac waveguide loop connects to outside waveguide
via Beam Splitter.
 On the waveguide loop,
a phase shifter can be used to adjusted the relative phase between the clockwise and counterclockwise
pulses;
(b) A five-level multi-$\Lambda$ system. Three excited states
$|1\rangle$, $|2\rangle$ and $|3\rangle$; Two ground states $|0\rangle$ and $|e\rangle$.
(c) A three-level multi-$\Lambda$ system. One excited state ($|1\rangle$) and two
ground states ($|0\rangle$ and $|e\rangle$).
}
\end{figure}
where $\sigma_{mn}=|m\rangle\langle n|(m,n=0, 1, 2, 3, e)$ are the transition operators
between the quantum state $|m\rangle$ and $|n\rangle$. $a_{jL}^\dagger(a_{jR}^\dagger)$ means the
creation operator for a left (right)-moving photon which drives the atomic transitions
$|j\rangle\Longleftrightarrow|0\rangle$.
The creation and annihilation operators of the waveguide
modes satisfy the commutation relation $[a_{jk}(\omega),a_{jk}^\dagger(\omega')]=\delta(\omega-\omega')$.
The corresponding frequency detuning is $\Delta_j=\omega_{j0}-\omega$ and $\omega_{j0}$ is the
frequency of transition $|j\rangle\Longleftrightarrow|0\rangle$.
When atom is initialed in the ground state $|0\rangle$ and the incoming photon only
couples to the atomic transition $|0\rangle\Longleftrightarrow|1\rangle$, we
can describe the evolution of the system under consideration by the
following wave function
\begin{equation}
 |\Psi(t)\rangle=\int_{-\infty}^{+\infty}\sum_{j=1,2,3}^{k=L,R}c_{jk}a_{jk}^\dagger(\omega)|0\rangle|vac\rangle d\omega
 +\sum_{j=1,2,3,e}c_j|j\rangle|vac\rangle,
 \end{equation}
where $|vac\rangle$ is zero photon Fock state of the waveguide mode and
$c_j (j=1, 2, 3, e)$ is the excitation amplitude of the atom in the state
$|j\rangle (j=1, 2, 3, e)$. $c_{jk} (j=1, 2, 3)$ is the amplitude for left
$(k=L)$- or right $(k=R)$-moving photon of the waveguide mode when
the atom is in the state $|0\rangle$. Using the Schr\"{o}dinger equation
$i\frac{\partial|\psi(t)\rangle}{\partial t}=H_I|\psi(t)\rangle$, we can obtain
the time evolution
of the amplitudes of the waveguide modes and atomic states are
\begin{subequations}
\begin{align}
i\dot{c}_{jk}&=\Delta_jc_{jk}-g_jc_j,
\\
i\dot{c}_{j}&=-g_j(c_{jk}+c_{j\bar{k}})-\Omega_jc_e-i\gamma_j/2,
\\
i\dot{c}_{e}&=-\sum_{j=1, 2, 3}\Omega_jc_j,
\end{align}
\end{subequations}
where $j\in\{1, 2, 3\}$, and $\{k, \bar{k}\}\in\{L, R\}$ with $k\neq\bar{k}$.
$\gamma_j$ is the
decay rate of the excited state $|j\rangle (j=1, 2, 3)$.

In this optical system, the spontaneous decay through the path
$|e\rangle\Longrightarrow|1\rangle\Longrightarrow|0\rangle$ can be greatly suppressed
by the relative phase $(\Theta)$. When the relative phase is tuned to be
zero, the incoming pulses propagating along opposite directions will form
the spatial superposition and constructively interfere at the position
of atom. Thus, combining with inherent symmetry of
Sagnac waveguide loop, the physical system under consideration can be described by the following
new operators:
\begin{subequations}
\begin{align}
a_{je(o)}(\omega)=\frac{1}{\sqrt{2}}[a_{jL}(\omega)\pm a_{jR}(\omega)],
\\
a_{je(o)}^\dagger(\omega)=\frac{1}{\sqrt{2}}[a_{jL}^\dagger(\omega)\pm a_{jR}^\dagger(\omega)],
\end{align}
\end{subequations}
where $a_{je(o)}(\omega)[a_{je(o)}^\dagger(\omega)](j=1, 2, 3)$ is the annihilation (creation)
operator of the waveguide and the subscript $e(o)$ denotes the even (odd) waveguide mode in
the Sagnac loop. In this case, the interaction Hamiltonian (1) can be rewritten as
\begin{align}
H_I=\int_{-\infty}^{+\infty}\sum_{j=1,2,3}[\Delta_ja_{je}^\dagger(\omega) a_{je}(\omega)+\Delta_ja_{jo}^\dagger(\omega) a_{jo}(\omega)
-\sqrt{2}g_ja_{je}^\dagger(\omega)\sigma_{0j}-\sqrt{2}g_ja_{je}(\omega)\sigma_{j0}]d\omega
-\sum_{j=1,2,3}\Omega_j(\sigma_{je}+\sigma_{ej}),
\end{align}
and the wave function (2) is
\begin{equation}
 |\Psi(t)\rangle=\int_{-\infty}^{+\infty}\sum_{j=1,2,3}^{k=e,o}c_{jk}a_{jk}^\dagger(\omega)|0\rangle|vac\rangle d\omega
 +\sum_{j=1,2,3,e}c_j|j\rangle|vac\rangle,
 \end{equation}
where
\begin{subequations}
\begin{align}
c_{je(o)}(\omega)&=\frac{1}{\sqrt{2}}[c_{jL}(\omega)\pm c_{jR}(\omega)],
\\
c_{je(o)}^\dagger(\omega)&=\frac{1}{\sqrt{2}}[c_{jL}^\dagger(\omega)\pm c_{jR}^\dagger(\omega)], j=1,2,3.
\end{align}
\end{subequations}
The above Hamiltonian (5) reveals that only the even mode experiences frequency conversion and
the frequency of odd mode remains unchanged. In fact,
when $\Theta=0$, only the even waveguide mode propagates in
the Sagnac loop \cite{prl12-108,pra12-85}.
Integrating the differential Eq. 3(a), and
substituting $c_{jk}$ into Eq. 3(b), we can obtain the coupling equations
connecting the input and output pulses
\begin{subequations}
\begin{align}
\dot{c}_{j}&=-2\pi g_j^2c_j+i2\sqrt{\pi}g_jc_{je}^{in}+i\Omega_jc_e, (j=1, 2, 3)
\\
\dot{c}_{j}&=2\pi g_j^2c_j+i2\sqrt{\pi}g_jc_{je}^{out}+i\Omega_jc_e, (j=1, 2, 3)
\end{align}
\end{subequations}
where $c_{je}^{out(in)}=\frac{1}{\sqrt{2}}(c_{jL}^{out(in)}+c_{jR}^{out(in)})$
($c_{jk}^{in}=\int_{-\infty}^{+\infty}c_{jk}(0)e^{-i\Delta_j(t-t_0)}d\Delta_j$ and
$c_{jk}^{out}=\int_{-\infty}^{+\infty}c_{jk}(0)e^{-i\Delta_j(t-t_1)}d\Delta_j$ ($j=1, 2, 3$ and $k=L, R$)
 correspond to the input and output waveguide modes, respectively.). Here, we have set the condition $t_0<t<t_1$.
$c_{jk}(0)$ is the initial value of the amplitudes $c_{jk}$ at $t=t_0$.
According to Eqs. 8(a)-(b), we can obtain the input-output formalism
\cite{book1,book2} as
\begin{equation}
c_{je}^{out}-c_{je}^{in}=i2\sqrt{\pi}g_jc_j, (j=1, 2, 3).
\end{equation}
Performing the Fourier transformations
$F(\Delta_1)=\frac{1}{\sqrt{2\pi}}\int_{-\infty}^{+\infty}f(t)e^{i\Delta_1t}d\Delta_1$ on
Eq. 3(c) and Eq. 8(a), substituting $c_j$ into Eq. (9),
and using the scattering matrix of the Sagnac interferometer \cite{jpb01-39,iee03-15}
 when $\Omega_3=0$, we can get the transport properties as
\begin{subequations}
\begin{align}
T_1(\Delta_1)&=\frac{c_{1e}^{out}}{c_{1e}^{in}}=1-\frac{2\Gamma_1(i\Delta_1A_2+\Omega_2^2)}{A_1(i\Delta_1A_2+\Omega_2^2)+\Omega_1^2A_2},
\\
T_2(\Delta_1)&=\frac{c_{2e}^{out}}{c_{2e}^{in}}=\frac{2\sqrt{\Gamma_1\Gamma_2}\Omega_1\Omega_2}{A_1(i\Delta_1A_2+\Omega_2^2)+\Omega_1^2A_2},
\\
T_3(\Delta_1)&=\frac{c_{3e}^{out}}{c_{3e}^{in}}=0,
\end{align}
\end{subequations}
where $A_j=i\Delta_1+\Gamma_j+\gamma_j/2$ and $\Gamma_j=2\pi g_j^2 (j=1, 2, 3)$.
For the case $\Delta_1=0$ and $\Omega_j\neq0(j=1,2,3)$, we have
\begin{subequations}
\begin{align}
T_1(\Delta_1)&=1-\frac{2\Gamma_1(\Gamma_2^{\prime}\Omega_3^2+\Gamma_3^{\prime}\Omega_2^2)}
{\Gamma_2^{\prime}\Gamma_3^{\prime}\Omega_1^2+\Gamma_1^{\prime}\Gamma_2^{\prime}\Omega_3^2+\Gamma_1^{\prime}\Gamma_3^{\prime}\Omega_2^2},
\\
T_2(\Delta_1)&=\frac{2\sqrt{\Gamma_1\Gamma_2}\Omega_1\Omega_2\Gamma_3^{\prime}}
{\Gamma_2^{\prime}\Gamma_3^{\prime}\Omega_1^2+\Gamma_1^{\prime}\Gamma_2^{\prime}\Omega_3^2+\Gamma_1^{\prime}\Gamma_3^{\prime}\Omega_2^2},
\\
T_3(\Delta_1)&=\frac{2\sqrt{\Gamma_1\Gamma_3}\Omega_1\Omega_3\Gamma_2^{\prime}}
{\Gamma_2^{\prime}\Gamma_3^{\prime}\Omega_1^2+\Gamma_1^{\prime}\Gamma_2^{\prime}\Omega_3^2+\Gamma_1^{\prime}\Gamma_3^{\prime}\Omega_2^2},
\end{align}
\end{subequations}
where $\Gamma_j^{\prime}=\Gamma_j+\gamma_j/2(j=1,2,3)$. The transmission coefficient $T_j(\Delta_1)$
 corresponds to the $j$th output pulse.
In the present scheme, selective frequency conversion is based on the selection of classical cw fields.
As shown in Fig. 1(b),
for instance, when the cw $\Omega_j(j=1, 2, 3)$ is cutoff, the output quantum filed $g_j$ is also
forbidden. The condition $\Omega_3=0$ leads to Eqs. 10, which indicates that the frequency of
the input quantum field $g_1$ may be converted from $\omega_{10}-\Delta_1$ into
$\omega_{20}-\Delta_2$ through adjusting parameters.

\begin{figure}[here]
\includegraphics[width=0.4\textwidth]{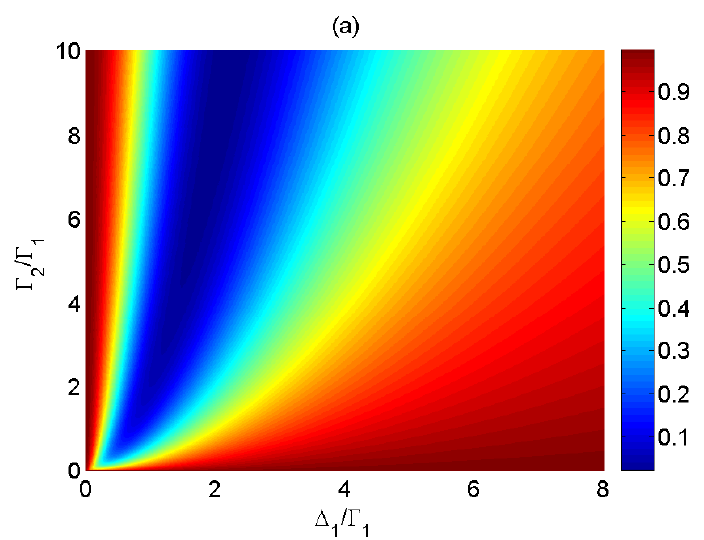}
\includegraphics[width=0.4\textwidth]{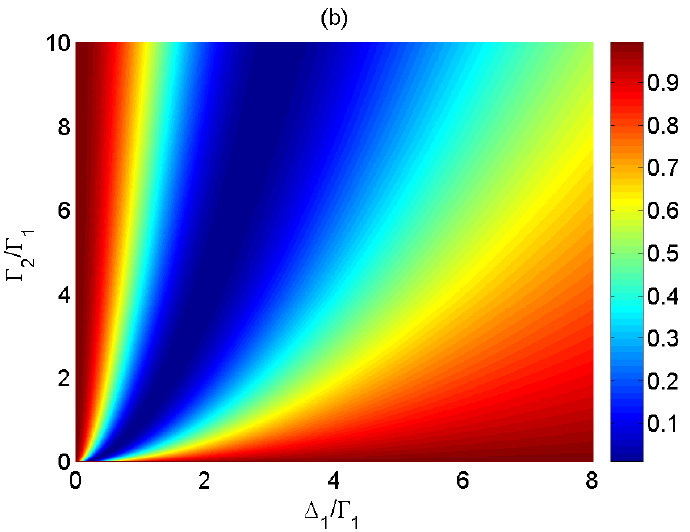}
\includegraphics[width=0.38\textwidth]{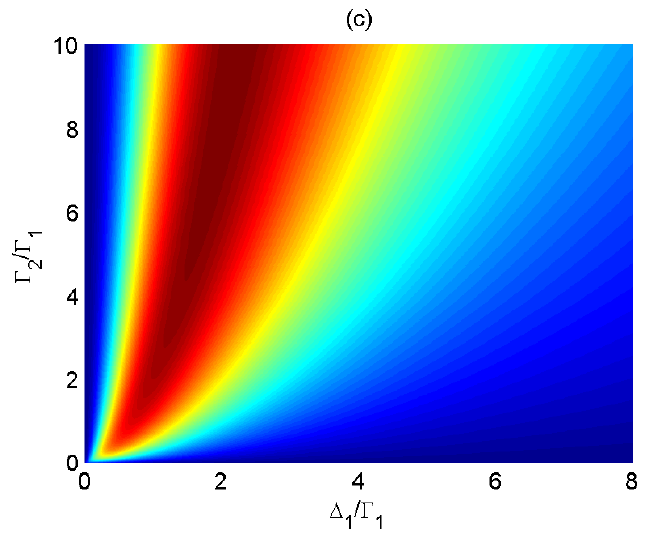}
\includegraphics[width=0.4\textwidth]{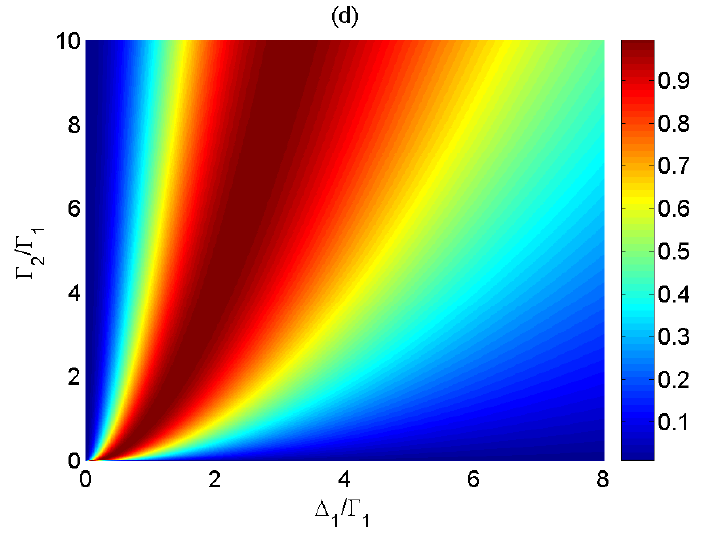}
\caption{\label{fig:fig4}(Color online) The transport properties
 versus $\Gamma_2/\Gamma_1$ and $\Delta_1/\Gamma_1$.
 $|T_1(\Delta_1)|^2$ with (a)
$\Omega_1=\Omega_2/\sqrt{2}=\Delta_1$
and (b)
$\Omega_1=\Omega_2=\Delta_1$;
$|T_2(\Delta_1)|^2$ with (c)
$\Omega_1=\Omega_2/\sqrt{2}=\Delta_1$
and (d)
$\Omega_1=\Omega_2=\Delta_1$.}
\end{figure}
\begin{figure}[here]
\includegraphics[width=0.4\textwidth]{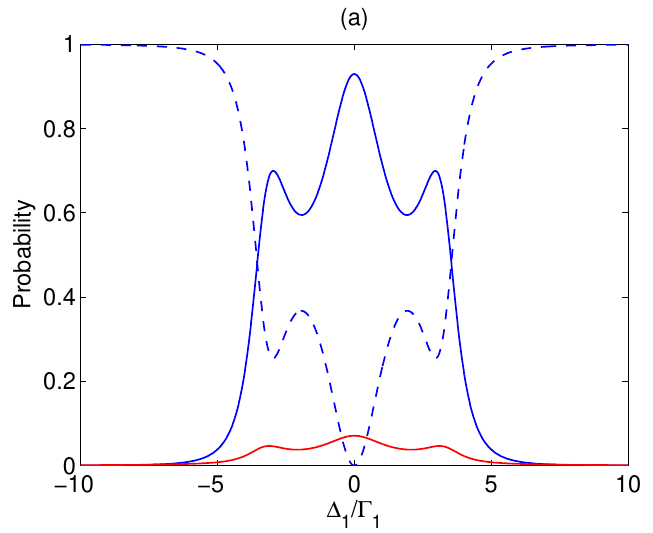}
\includegraphics[width=0.4\textwidth]{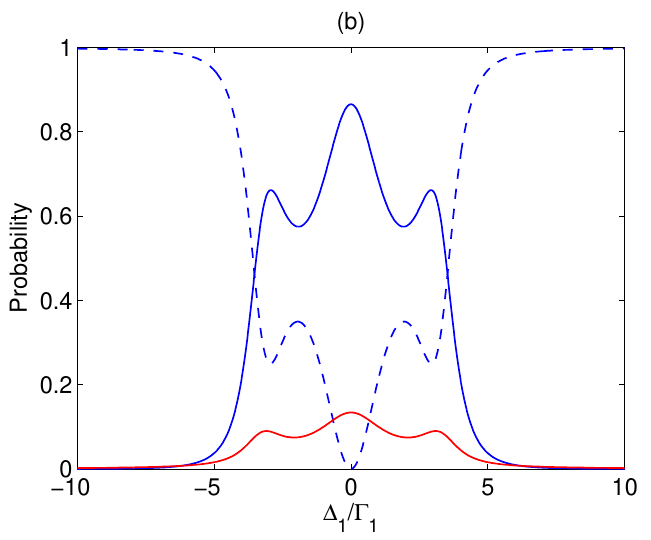}
\includegraphics[width=0.4\textwidth]{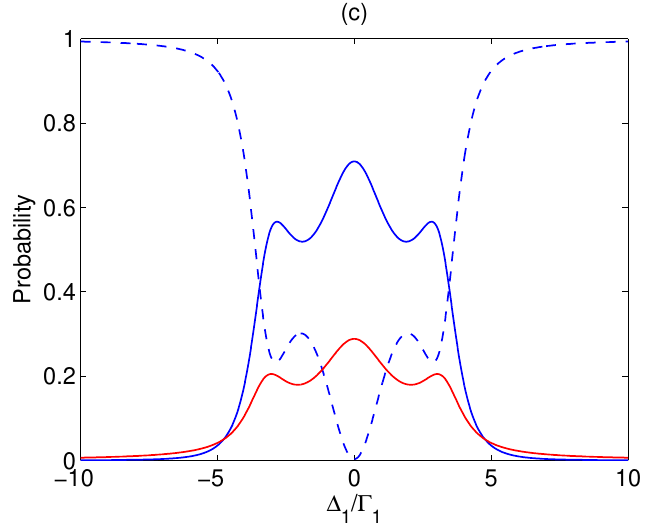}
\includegraphics[width=0.4\textwidth]{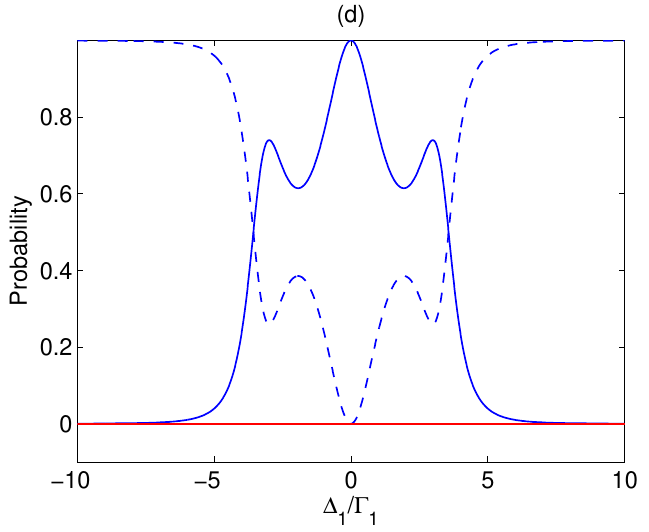}
\caption{\label{fig:fig4}(Color online)
Probability
$|T_1(\Delta_1)|^2$ (blue dash lines), $|T_2(\Delta_1)|^2$ (blue solid lines) and
loss probability of photon $P_{loss}=1-|T_1|^2-|T_2|^2$
 (red solid lines) versus frequency detuning $\Delta_1/\Gamma_1$ with (a) $\gamma_1=0.1\Gamma_1$; (b)$\gamma_1=0.2\Gamma_1$;
 (c) $\gamma_1=0.5\Gamma_1$; (d)$\gamma_1=0$;
 Other parameters are chosen as $\Gamma_2=\Omega_1=\Omega_2/\sqrt{2}=2\Gamma_1$ and $\gamma_2\simeq\gamma_1$.}
\end{figure}
\section{Analysis and discussion}
For simplicity, we consider first the case of $\gamma_j=0$.
By studying Eqs. 10(a)-(b), we can try to find the appropriate parameters
which lead to perfect frequency conversion. When parameters satisfy the following
conditions
\begin{subequations}
\begin{align}
(\Omega_1^2-\Delta_1^2)\Gamma_2+(\Delta_1^2-\Omega_2^2)=0\\
\Delta_1(\Omega_1^2+\Omega_2^2-\Gamma_1\Gamma_2-\Delta_1^2)=0,
\end{align}
\end{subequations}
the efficiency of quantum frequency conversion reaches unity, i.e., $|T_1(\Delta_1)|^2=0$ and
$|T_2(\Delta_1)|^2=1$. It is easy to find that we can obtain wonderful frequency conversion
not only for resonant case but also for off-resonant case. For example, the condition
$\Omega_1^2\Gamma_2=\Omega_2^2\Gamma_1$ ensures the complete frequency
conversion for the resonant case $\Delta_1=0$. More generally, we further
described the effect of parameters on the transport properties in Fig. 2(a)-(d),
which show that the efficiency near unity is achievable with
 frequency detuning and coupling strengths $\Gamma_j(j=1,2)$ in a wide region.

 Spontaneous deacy of the atomic excited states will seriously influence the efficiency of
 quantum frequency conversion.
As shown in Fig. 3, the converted probability of photon $|T_2(\Delta_1)|^2$
declines dramatically as the decay rates increase. In the meantime
the loss probability of photon $P_{loss}$ increases sharply.
Nevertheless, even then we can get the converted probability $|T_2(\Delta_1)|^2$
more than $86\%$ with $\gamma_1\simeq\gamma_2=0.2\Gamma_1$.
 In addition, the converted probability $|T_2(\Delta_1)|^2$
 and loss probability $P_{loss}$ reach maximum at the resonant point with the parameters
 $\Gamma_2=\Omega_1=\Omega_2/\sqrt{2}=2\Gamma_1$, which just satisfy the above condition
 $\Omega_1^2\Gamma_2=\Omega_2^2\Gamma_1$ for the resonant case.
 The probability $|T_1(\Delta_1)|^2$, however, is not obviously affected
 by the dissipation, and is nearly zero at resonant point for different
 decay rate.

The present scheme under consideration is capable of generating mode entanglement of photons.
From Eqs. 10, we can find that the maximum mode entanglement of two photons can be realized by
adjusting parameters until
$|T_1(\omega)|^2=|T_2(\omega)|^2=1/2$. In the following discussion, we focus on
generation of the $W$ entangled state of photons of different frequencies.
For simplicity, we consider only the resonant case, i.e., $\Delta_1=0$ and obtain
the transport properties as described in Eqs. 11. The parameters conditions
$\Gamma_2=\Gamma_3$, $\Omega_2=\Omega_3$ and $\Omega_2/\Omega_1=(\sqrt{3}-1)\sqrt{\Gamma_2/\Gamma_1}/2$
will lead to $|T_1(\omega)|^2=|T_2(\omega)|^2=|T_3(\omega)|^2=1/3$.
In this case, the output pulses are in the standard $W$ entangled state, i.e.,
$|\psi\rangle_w=\frac{1}{\sqrt{3}}(|1_{\omega_{10}}00\rangle+|01_{\omega_{20}}0\rangle+|001_{\omega_{30}}\rangle)$.
Here, the states $|1_{\omega_{10}}00\rangle$, $|01_{\omega_{20}}0\rangle$ and $|001_{\omega_{30}}\rangle)$ correspond
to transport pulses $|T_1(\omega)|^2$, $|T_2(\omega)|^2$ and $|T_3(\omega)|^2$, respectively.
In fact, by choosing intensities of cw
properly, one can obtain different transport properties $|T_j(\omega)|^2(j=1,2,3)$.
To be clear, we describe the changes of transport properties $|T_j(\omega)|^2(j=1,2,3)$ and fidelity
of entangled states with
different experimental parameters in Fig. 4(a)-(b), which show the present scheme has
flexible adjustment of parameters. In addition, we consider the effect of
decay rate on the fidelity of W entangled state in Fig. 4(c). When $\gamma/\Gamma_1=0.5$,
the fidelity $F$ of the W of photons of three kinds of frequencies is more than $91\%$.
\begin{figure}[here]
\includegraphics[width=0.30\textwidth]{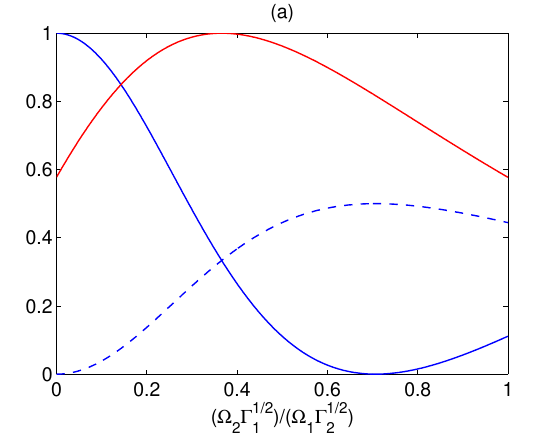}
\includegraphics[width=0.32\textwidth]{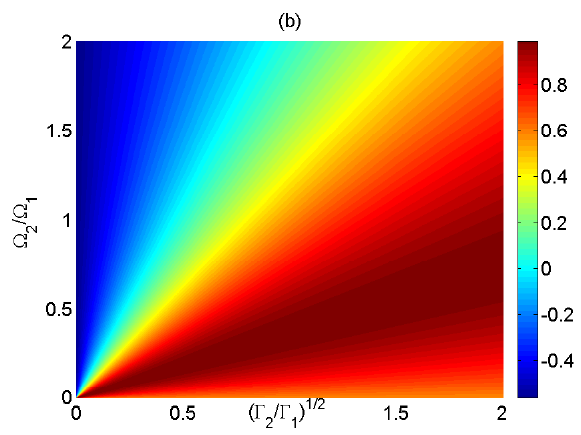}
\includegraphics[width=0.3\textwidth]{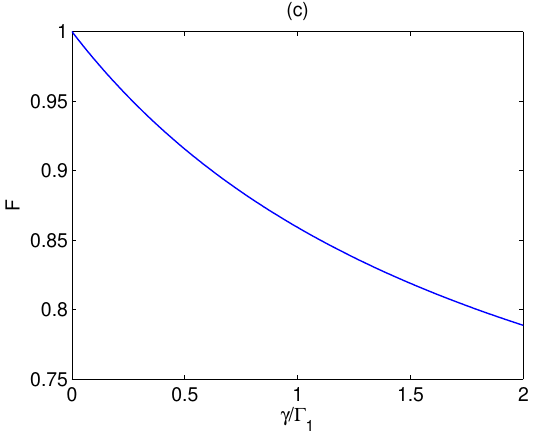}
\caption{\label{fig:fig4}(Color online) (a)The fidelity of photonic entanglement $F$ (red solid lines)
and the probability $|T_1(\omega)|^2$ (blue solid lines), $|T_2(\omega)|^2=T_3(\omega)|^2$ (blue dash lines)
as functions of $(\Omega_2\Gamma_1^{1/2})/(\Omega_1\Gamma_2^{1/2})$ and
$\Delta_1/\Gamma_1$ with $\Omega_2=2(2-\sqrt{2})\Gamma_1$;
(b)The fidelity $F$ of the $W$ entanglement of three frequencies modes against
$\Omega_2/\Gamma_1$ and $\Gamma_2^{1/2}/\Gamma_1^{1/2}$; (c)The fidelity $F$ against
decay rate $\gamma/\Gamma_1$ with $\gamma\simeq\gamma_j(j=1,2,3)$ and
$\Omega_2/\Omega_1=(\sqrt{3}-1)\sqrt{\Gamma_2/\Gamma_1}/2$; Other
parameters are $\Omega_2\simeq\Omega_3$ and $\Gamma_2\simeq\Gamma_3$.}
\end{figure}

In the above analysis, we have carried out extensive numerical calculations to establish
the validity of the present scheme by experimentally achievable parameters.
Cold $^{87}$Rb atoms can be chosen as a possible experimental candidate system for the present
scheme. We take, for example,  $|0\rangle=|5S_{1/2}, F=1\rangle$,
$|e\rangle=|5S_{1/2}, F=2\rangle$, $|1\rangle=|5P_{3/2}, F=0\rangle$, $|2\rangle=|5P_{3/2}, F=1\rangle$ and
$|3\rangle=|5P_{3/2}, F=2\rangle$, which correspond to the multi-$\Lambda$ configuration shown in Fig. 1(b).
 The appropriate transitions are
$|0\rangle\Longleftrightarrow|j\rangle$ and $|e\rangle\Longleftrightarrow|j\rangle(j=1,2,3)$
with $\gamma_{je}=\gamma_{j1}=\gamma_j/2=2\pi\times6$ MHz.
Notice that some works are based on a three-level and multi-$\Lambda$ system, including
four-wave mixing \cite{olwu04-29,wu11} and induced transparency \cite{pd}. As shown in
Fig. 1(c), a three-state and multi-$\Lambda$ system is suitable for the present scheme.
We can also get the same result through the process similar to the above.

To summary, we propose a theoretical scheme for selective quantum frequency conversion and multifrequency mode
entanglement via input-output formalism. The frequency of the output photon can be selected by applying
proper classical fields. The probability amplitudes of the output photons also can be adjusted by
 choosing intensities of driving fields and other experimental parameters properly. The adjustable
 probability amplitudes  will
 lead to the generation of multifrequency mode $W$ entanglement.
Even for the existence of spontaneous decay of excited states, we can obtain the highly efficient
frequency conversion and multifrequency mode $W$ entanglement with high fidelity.
In addition, our calculations
show that the present scheme can work well in a wide parameters region.

\section*{Acknowledgments}
The work is supported in part by National Basic Research Program of China (No. 2012CB922103) and by the National
Science Foundation (NSF) of China (Grant Nos. 10874050, and 11005057) and the Fundamental Research Funds for the
Central Universities and by the Scientific and Technological Research Program of Education Department of Hubei Province (No.
Z200722001 and No.B20122201). The authors acknowledge Prof. Ying Wu for his enlightening suggestions.

\end{document}